\documentclass[submission,copyright,creativecommons]{eptcs} 
\providecommand{\event}{ICLP 2021}
\usepackage{caption} 
\usepackage[fleqn,alignedleftspaceno]{amsmath}
\usepackage{amssymb}
\usepackage{mathrsfs}
\usepackage{tikz}
\usepackage{tikz-qtree}
\usepackage{multirow}
\usepackage{soul}
\usepackage{listings}
\usepackage{algorithm}
\usepackage{algpseudocode}
\usepackage{algorithmicx}
\usepackage[english]{babel}
\usepackage{biblatex}
\usepackage{booktabs}
\usepackage{csquotes}
\usepackage[linewidth=0.5pt]{mdframed}
\usetikzlibrary{positioning}
\usetikzlibrary{trees}
\usetikzlibrary{decorations.pathreplacing,calc}
\usetikzlibrary{shapes.misc}
\usetikzlibrary{calc,shapes.multipart,chains,arrows}

\newcommand{\mycomment}[1]{}
 \usepackage[compact]{titlesec}
 \titlespacing{\section}{1pt}{1pt}{1pt}
 \titlespacing{\subsection}{2pt}{2pt}{2pt}

\def\authorrunning{S.C. Varanasi, N. Mittal \& G. Gupta}
\def\titlerunning{Concurrent Data Structure Synthesis using ASP} 
\begin{document}

\title{Generating Concurrent Programs  from Sequential Data Structure Knowledge  using Answer Set Programming
}
\author{
    % Authors
        Sarat Chandra Varanasi \and 
        Neeraj Mittal \and
        Gopal Gupta  
    \institute{      Department of Computer Science, 
    The University of Texas at Dallas, 
     Richardson, Texas, USA}
    \email{\{sarat-chandra.varanasi, 
    neerajm, gupta\}@utdallas.edu}
}

\maketitle

\begin{abstract}
 We tackle the problem of automatically designing concurrent data structure operations given a sequential data structure specification and knowledge about concurrent behavior.  Designing concurrent code is a non-trivial task even in simplest of cases.  Humans often design concurrent data structure operations by transforming sequential versions into their respective concurrent versions.  This requires an understanding of the data structure, its sequential behavior, thread interactions during concurrent execution and shared memory synchronization primitives. We mechanize this design process using automated commonsense reasoning. 
We assume that the data structure description is provided as axioms alongside the sequential code of its algebraic operations. This information is used to automatically derive concurrent code for that data structure, such as dictionary operations for linked lists and binary search trees. Knowledge in our case is expressed using Answer Set Programming (ASP), and we employ deduction and abduction---just as humans do---in the reasoning involved.
ASP allows for succinct modeling of first order theories of pointer data structures, run-time thread interactions and shared memory synchronization.  Our reasoner can systematically make the same judgments as a human reasoner, while constructing provably safe concurrent code. We present several reasoning challenges involved in transforming the sequential data structure into its equivalent concurrent version. All the reasoning tasks are encoded in ASP and our reasoner can make sound judgments to transform sequential code into concurrent code. To the best of our knowledge, our work is the first one to use commonsense reasoning to automatically transform sequential programs into concurrent code. We also have developed a tool that we describe that relies on state-of-the-art ASP solvers and performs the reasoning tasks involved to generate concurrent code.
\end{abstract}

\section{Introduction}
    We present a novel technique based on answer set programming that automatically generates concurrent programs for pointer data structures given a first order (logic) data structure theory, background knowledge about its sequential operations and axioms for concurrency. Design of concurrent operations for data structures is non-trivial. There are several challenges that need to be addressed for automatically generating concurrent algorithms. Traditionally, concurrent programs are designed manually and their proofs of correctness are done by hand. Few concurrent data structures are also verified using symbolic bounded model checking \cite{vechev2010abstraction}. Avoiding state space explosion in the verification of concurrent programs is the main challenge for symbolic model checkers. Several works address this issue in interesting ways \cite{emerson2000reducing, vechev2008deriving}. Other formal approaches involve performing Hoare-Style Rely-Guarantee reasoning \cite{vafeiadis2006proving} to verify concurrent programs that have been manually designed. 
    %GG:
    These approaches, thus, seek help of automated verification in an otherwise manual design process to ensure correctness. In contrast, our approach leverages reasoning techniques employed in AI,  knowledge about concurrency, and explicitly-modeled sequential data structure code to automatically derive a safe concurrent program. Work in model checking and formal logics for concurrency do not exploit the sequential data structure knowledge. Their main focus is to prove absence of incorrect thread interactions (or traces). The proof of correctness of the verified concurrent code is provided outside their frameworks, assuming certain symmetry properties on concurrent interactions. Our work, in contrast, performs the reasoning tasks that an expert in concurrent program design performs in order to construct a safe concurrent program. This requires an understanding of the data structure representation, the library of algebraic operations that modify the data structure, an understanding of shared memory and how primitive read and write operations affect the shared memory. Additionally, the expert can explicitly describe the safety conditions that are desired, the invariants that need to be preserved during concurrent execution. With this knowledge, the expert creates the concurrent program that acquires the ``right" number of locks (synchronization steps) that is safe for any concurrent interaction with an unbounded number of threads. The work reported here aims to automate this process, thereby deriving concurrent code automatically.
    
    Our work can be seen as applying automated (commonsense) reasoning \cite{mccarthy1960programs} to the program synthesis problem. To the best of our knowledge our is the first effort that attempts to emulate the mind of a human domain expert who designs concurrent data structure using sequential ones as a starting point. Our goal is that once axioms for concurrency, etc., have been developed by a domain expert, a programmer can simply specify a new sequential data-structure and automatically obtain its correct concurrent version.
    
    We make extensive use of answer set programming (ASP) to model the domain expert's thought process.  We assume basic familiarity with stable model semantics and ASP solvers \cite{gebser2016theory}. More details about ASP can be found elsewhere \cite{gelfond1988stable,gelfond2014knowledge}.
    We next give a background of various notions associated with Data Structures and Concurrency in Section 2. It also provides an example of a Data Structure definition and its sequential program knowledge. We illustrate how this knowledge is used to translate the sequential code into concurrent code. The various reasoning steps involved in the transformation are shown explicitly in a side-by-side comparison. Section 3 introduces general challenges involved in transforming a sequential data structure into its concurrent version. Section 4 introduces our technique and explains thread interference, predicate falsification and the role of locks to preserve data structure invariants. Section 5 discharges the ideas into several theories involved in making the reasoning tasks executable. These theories are ASP programs that check predicate falsification and infer thread synchronization (using locks). They also infer the correct order of program execution needed in a concurrent program in order to preserve data structure invariants. Section 6 provides the soundness proof of our approach. Section 7 gives details of our implemented tool: \textit{Locksynth}. Section 8 concludes with our Experimental setup and future work. 
\section{Background and General Notions}
  
  \mycomment{
  ocksynth\. subsection{Answer Set Programming}
    Answer Set Programming is a declarative problem solving paradigm with applications spanning several areas of AI research: from planning to complex human style commonsense reasoning \cite{erdem2016applications,chen2016physician}. The expressive power of ASP is due to its non-monotonic reasoning capabilities. Non-monotonic reasoning allows one to retract conclusions in light of new evidence. 
    Fundamentally, ASP programs are normal logic programs that are non-monotonic, i.e., we can write code in ASP to take an action if a proof fails. This is achieved through support for negation as failure. This helps model commonsense reasoning (humans can take an action predicated on failure of a proof). Monotonic logics cannot reason about proof-failure within the logic itself. 
    
    An ASP Program consists of rules of the form $\{p \leftarrow q_1, q_2,..q_i, not \ r_1, not \ r_2, .., \ not \ r_j\}$. If $i = j = 0$, then $p$ is a fact. If $p$ is the empty ($\square$ or false), the rule represents a constraint. The operator $not$ represents negation-as-failure. The set of satisfiable literals of an answer set program are termed as its answer sets (stable models). }
    %An atom $m$ is entailed by an ASP program $\Pi$, ie., $\Pi \models m$ if and only if $m$ is present in every stable model of $\Pi$ \cite{faber2009manifold}. 
     \subsection{Data Structures}
   
    \noindent     Data structures include some representation of information and the dictionary operations associated with them such as membership, insert and delete. Representation itself involves several notions at various levels of abstraction. For example, to describe a linked list, one needs primitive notions of \textit{nodes} contained in memory, connected by a chain of \textit{edges}. Further, there are notions of \textit{reachability} (or \textit{unreachability}) of nodes and keys being present (or \textit{absent}) in a list. Membership operations usually involves traversing the elements (or nodes) in the data structure until an element(s) satisfying certain criteria is found. Insert operation (and similarly delete operation) also involves traversing the data structure until a right ``window" of insertion is found. The notion of window represents some local fragment of the data structure that is modified as part of a data structure update operation. This notion is useful when discussing about locking nodes in concurrent programs.
    %\subsection{Tree-Based Pointer Data structures} 
    \\ \\ 
    \par\noindent\textbf{\textit{Tree-Based Pointer Data structures}}
     A heap is a collection of nodes connected by edges. A data structure $\mathcal{D}$ is a recursive definition defining a tree of nodes in memory. The recursive definition for $\mathcal{D}$ can be viewed as a constructor of various instances of the data structure. Further, the only primitive destructive operation that may be performed is linkage of pointers: $link(x,y)$. The abstract relation $link(x,y)$ links node $y$ to $x$ in the heap. We support transformation of concurrent code for an algebraic operation $\sigma_{\mathcal{D}}$ associated with $\mathcal{D}$ such that $\sigma_{\mathcal{D}}$ may be performed in a constant number of $link$ operations. For instance insert operation for linked lists can be performed in two steps. 
  \subsection{Concurrent Data Structures}
    Concurrent Data Structures usually support data structure dictionary operations being manipulated by an unbounded number of interacting threads. They are nothing but multiprocessor programs.  We only assume a sequentially consistent shared memory model in this paper. Sequential consistent memory allows any update performed on the shared memory to be visible, before performing a subsequent read, to every thread in the system. Concurrency can be viewed as a sequence of interleaved steps taken by various threads in the system. 
    %A concurrent program is the set of interleaved traces it generates. 
     To  make sense of correctness of concurrent data structures, the notion of \textit{linearizability} \cite{herlihy2011art} is widely used. A concurrent data structure is termed linearizable, if the effects of concurrent modification by several threads can be viewed as if the concurrent operations were performed in some sequential order. In this paper, we study the modifications performed on a data structure as if they are respecting a given sequential algorithm. This allows us to model concurrency in an intuitive manner and sidesteps the necessity to understand traces. This assumption is sufficient to generate safe concurrent programs.  
     %do we need to worry about deadlock-freedom? omit the sentence?
     %sv: omitted

    \subsection{Data Structure Theory and Knowledge}
  We assume that a first order theory $\mathcal{T}$ is provided for a pointer data structure along with the sequential data structure knowledge $\mathcal{K}$ (see Fig. \ref{figure:fig1}). We use the theory for linked lists and its knowledge as running example in this paper. The technique however applies to all tree-based pointer data structures. The data structure theory defines linked lists as a chain of edges with special sentinel nodes $h$ at the head of the list and $t$ at the end. The meaning of predicates $reach$ and $present$ is straightforward.
  
  \subsection{Sequential Data Structure Knowledge}
  %\par\noindent\textbf{\textit{Sequential Data Structure Knowledge}}
  The knowledge $\mathcal{K}$ contains which predicates are time-dependent (fluents). It also has the $start$, $end$ and $next$ nodes for a data structure traversal, beginning from the start node. It also contains the pre/post-conditions of insert and delete operations for linked lists. The primitive write step is denoted by \textit{link} (link-pointer) operation. The effects of link operation are also described using \textit{causes} relation. The knowledge $\mathcal{K}$ is useful for two purposes: 1. It bounds the interference effects of arbitrary thread interactions in a concurrent execution. 2. It narrows the code blocks that need to be synchronized to obtain a concurrent algorithm. However, as we present next, there are several challenges to transforms steps $\langle1,2\rangle$ of \textit{insert} operation (Fig. \ref{figure:fig2}) into a concurrent version.  
  The program statements are encoded within the vocabulary of the data structure using answer set programming (ASP). Program Blocks in computer programs can be viewed as equivalence class of input-output transformation.
   Further, the program blocks perform destructive update operations on the data-structure (insert/delete). We assume program blocks are straight line programs. If the sequential program has multiple blocks, the conditions under which the blocks may be executed should be mutually exclusive. In other words, every precondition uniquely determines the its associated program block. 
   Given the data structure definition, it is straightforward to generate data structure instances that satisfy a given equivalence class. This is because the assumed data structure definition $\mathcal{D}$ is recursive. The recursive definition for $\mathcal{D}$ can enumerate the set $S_{\mathcal{D}}$ of all structurally isomorphic instances of $\mathcal{D}$. The set $S_{\mathcal{D}}$ can be ordered by the number of recursion unfoldings used to generate the instances, starting from the least number of unfoldings. For $\mathcal{D}_i, \mathcal{D}_j \in S_{\mathcal{D}}$, $i < j$ implies that $\mathcal{D}_i$ is a ``smaller" structure than $\mathcal{D}_j$ and appears before $\mathcal{D}_j$ in the recursion depth ordering.  
  \begin{figure} 
   \fbox{\scriptsize
                  \begin{tabular}{p{6cm}|p{8.75cm}}
                            {\bf 
                             Original Theory $\mathcal{T}$}  \newline
                           {
                            \begin{minipage}{0.97\linewidth} 
                              \fbox{
                              \textit{List Structural Definition}} 
                               \begin{align*}
                                list   \leftarrow & ~edge(h, X), key(h, K_h), \\ 
                                       \phantom{\leftarrow} & ~key(X, K_X),   ~K_h <  K_X, \mathit{suffix(X)} 
                             \\
                         \mathit{suffix(t)} \leftarrow  &
                         \\
                             \mathit{suffix(X)}  \leftarrow  & ~edge(X, Y),  key(X, K_X), \\ 
                             \phantom{\leftarrow} & ~key(Y, K_Y), 
                                ~K_X < K_Y, \mathit{suffix(Y)} 
                             \end{align*}
                   
                             \fbox{\textit{Reachability Definition}} 
                          \begin{align*} 
                            reach(h) \leftarrow & \\
                            reach(X) \leftarrow & ~edge(Y, X), reach(Y) 
                            \end{align*} 
                            \fbox{\textit{Keys Present Definition}} 
                            \begin{align*}
                            present(K) \leftarrow & ~reach(X), key(X, K)
                            \end{align*} 
                            \textbf{Data Structure Knowledge} $\mathcal{K}$ 
                             \\
                        \fbox{\textit{Fluents, Start and End nodes}
                        }
                            \begin{align*}
                                start\_node(h) \leftarrow & \\
                                end\_node(t) \leftarrow &
                            \\ next\_node(x,y,\tau) \leftarrow & ~edge(x,y), \\ \phantom{\leftarrow} & ~key(x,k_x), lt(k_x,k_{\tau}) \\
                            invariant(list) \leftarrow & \\
                            \mathit{fluent(list)} \leftarrow & \\ 
                            \mathit{fluent(reach)} \leftarrow & \\
                            \mathit{fluent(suffix)} \leftarrow & \\
                            \mathit{fluent(edge)} \leftarrow & \\ 
                          \mathit{fluent(present)} \leftarrow &
                          \\ \mathit{fluent(next\_node)} 
                          \leftarrow &
                            \end{align*}
                          \end{minipage}
                          }
                           & 
                               \textbf{Data Structure Knowledge} $\mathcal{K}$ \newline
                           {  
                        \begin{minipage}{0.97\linewidth}
                             \fbox{\textit{Pre/Post Condition(s) (Insert)}} 
                           \\   \begin{align*}
                              pre(ins, block1,  [reach(x), \mathit{suffix(y)}, edge(x, y),  
                             key(x, k_x),  \\ key(y, k_y), key(\tau, k_\tau), k_x < k_{\tau}, k_{\tau} <  k_y])   & \leftarrow   \\
                             post(ins, [reach(\tau), edge(\tau, y), edge(x, \tau)]) & \leftarrow
                             \end{align*}
                             \fbox{\textit{Program Steps (Insert)}} 
                              \begin{align*}
                               step(1, ins, block1, link(x,\tau)) & \leftarrow\\
                               step(2, ins, block1, link(\tau, y)) & \leftarrow
                              \end{align*}        
                             \fbox{\textit{Primitive Destructive Update Step}} 
                              \begin{align*}
                                   primitive(link(x,y),\mathit{modifies(x)}) & \leftarrow \\
                                   causes(edge(x,y),link(x,y)) & \leftarrow 
                               \end{align*}
                                  \fbox{\textit{Pre/Post Condition(s) (Delete)}} 
                             \begin{align*}
                             pre(del, block1, [reach(x), \mathit{suffix(z)}, edge(x, y), edge(y, z), k_\tau = k_y]) & \leftarrow  \\
                                post(del, [not \ reach(y), edge(x, z]) & \leftarrow
                             \end{align*}
                             \fbox{\textit{Program Steps (Delete)}} 
                              \begin{align*}
                               step(1, del, block1, link(x,z)) & \leftarrow
                               \end{align*}
                        \end{minipage}
                        }
                            \end{tabular}
                           }
      \caption{Data Structure Theory and Knowledge}
      \label{figure:fig1}
   \end{figure}
   
\subsection{Illustration of lazy synchronization for Concurrent Linked List Insert}
  Consider the insert operation of a linked list. We insert a target node (key) that preserves the linked list order. The left hand side of Fig. \ref{figure:fig1}
  shows the sequential code. The code is annotated with necessary pre-condition and post-conditions. The $list$ invariant is violated after step 1 and restored back in step 2. The equivalent concurrent code is shown on the right. In a concurrent setting, $list$ cannot be violated at any time. This is because a membership operation cannot encounter a broken list. Therefore, order of the program steps matter in a concurrent execution. Also, the right locks have to be acquired to perform safe destructive updates. Figuring out the right nodes to be locked requires considerable domain expertise and understanding of the data structure. Further, acquiring the precise locks is still insufficient. As seen in the figure, there is an extra validation of the reachability of the nodes post lock acquisition. This is because, by the time lock is acquired on node $x$, the node may have been removed from the list. Hence, we do the reachability check. This technique is widely known as \textit{lazy synchronization} \cite{herlihy2011art}.  
  \begin{figure}
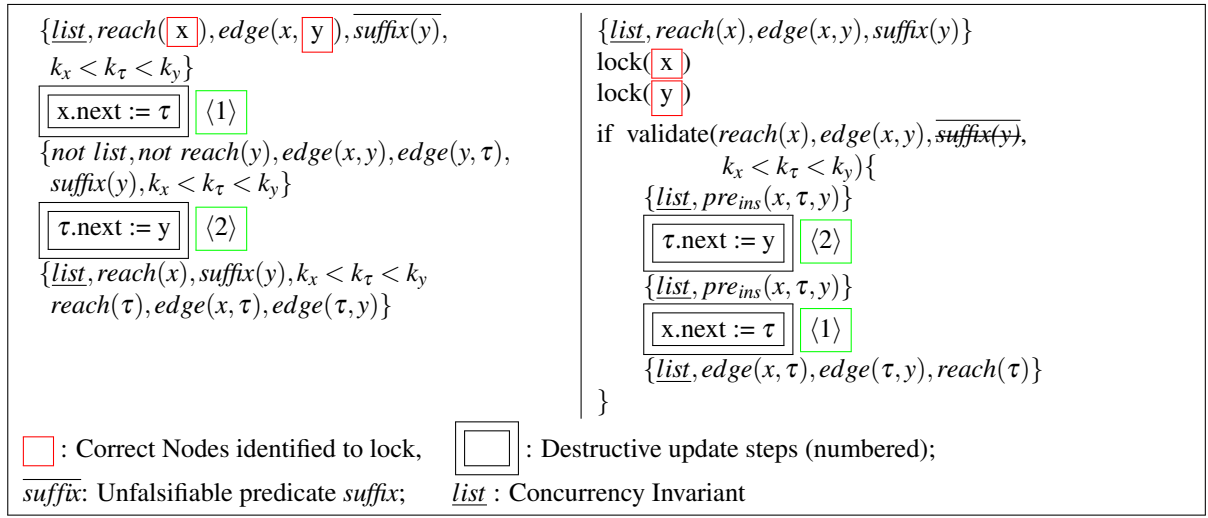

  \small
  \fbox{
  \begin{minipage}{0.97\linewidth}
      \begin{tabular}{p{7cm}|p{7cm}}
           $\{\underline{list}, reach(\fcolorbox{red}{white}{x}), edge(x,\fcolorbox{red}{white}{y}), \overline{\mathit{suffix(y)}},  \newline \phantom{x} k_x < k_{\tau} < k_y\}$ \newline
              \fbox{\fbox{x.next := $\tau$}}  \fcolorbox{green}{white}{$\langle1\rangle$}   \newline
            $\{not \ list, not \ reach(y), edge(x,y), edge(y,\tau),$ \newline
            $ \phantom{x} \mathit{suffix(y), k_x < k_{\tau} < k_y}\}$ \newline
              \fbox{\fbox{$\tau$.next := y}}  \fcolorbox{green}{white}{$\langle2\rangle$} \newline   
           $\{\underline{list}, reach(x),  \mathit{suffix(y), k_x < k_{\tau} < k_y}$ \newline
            $\phantom{x} reach(\tau), edge(x,\tau), edge(\tau,y)\}$ 
          & 
          $\{\underline{list}, reach(x), edge(x,y), \mathit{suffix(y)}\}$ \newline
             lock(\fcolorbox{red}{white}{x}) \newline               
             lock(\fcolorbox{red}{white}{y}) \newline
        %  $\{\underline{list}, pre\_ins?\footnote{has to validated post lock-acquisition}(x,\tau,y)\}$ \newline 
        if \ validate($reach(x),edge(x,y), \overline{\textit{\st{suffix(y)}}}$, \newline
            \phantom{xxxxxxxxx} $k_x < k_{\tau} < k_y)\{$  \newline 
          \phantom{xxx} $\{\underline{list}, pre_{ins}(x,\tau,y)\}$ \newline
          \phantom{xxx}    \fbox{\fbox{$\tau$.next := y}} \fcolorbox{green}{white}{$\langle2\rangle$} \newline
          \phantom{xxx} $\{\underline{list}, pre_{ins}(x,\tau,y)\}$ \newline
          \phantom{xxx}     \fbox{\fbox{x.next := $\tau$}} \fcolorbox{green}{white}{$\langle1\rangle$} \newline
          \phantom{xxx} $\{\underline{list}, edge(x,\tau), edge(\tau, y), reach(\tau)\}$ \newline
          $\}\phantom{thequickbrownfoxjumpedoverthelazydogthequick}$
      \end{tabular}
         \\
               \fcolorbox{red}{white}{\phantom{x}} : Correct Nodes identified to lock,  
         ~~~~\fbox{\fbox{\phantom{lin}}} : Destructive update steps (numbered); \\
         ~~~~~~~~$\overline{su\!f\!\!f\!i\!x}$: Unfalsifiable predicate $\mathit{suffix}$; 
         ~~~~~~$\underline{list}$ : Concurrency Invariant 
\end{minipage}
}
\caption{Steps Involved in Transforming Sequential Linked List to Lazy Concurrent List}
\label{figure:fig2}
\end{figure}

\section{Sequential Data Structures Code to Concurrent Code: Challenges}
     We assume that the traversal code remains the same as the sequential version for a lock-based concurrent data structure. 
     Therefore, the challenges we discuss are purely for destructive update program steps. We present the challenges involved and how they are addressed in turn. 
    \subsection{Order of the Program Steps Matter}
    
        We have already shown that the order of the program steps matter. However, 
        in general, it is possible that there exists no ordering of steps that preserves an invariant in a concurrent execution. For instance, the Internal BST invariant cannot be maintained by any order of the program steps involved in either inserting or deleting a node into the BST. In such a case, the designer uses the Read-Copy-Update (RCU) \cite{mckenney2020rcu} technique to copy the window, perform changes locally (outside shared memory) and atomically splice window back to the shared memory. The RCU technique depends on the ability to splice back the window atomically. For tree-data structures, if the window is a sub-tree, then it is easy to atomically splice a sub-tree to shared memory by updating its parent pointer, in the shared memory. The applicability of RCU framework can be either made explicit in the data structure knowledge, or should otherwise be inferrable from the knowledge of data structure representation/operations.   
\subsection{Concurrent Traversal may require RCU}
As we mentioned before, our assumption is that transformation for concurrent membership operations is vacuous (i.e., membership operation is unchanged in a concurrent setting). This ensures that the membership queries execute as fast as possible while acquiring no locks. However, for the membership operation to work consistently with insert/delete, the code for insert and delete operations should work correctly. We illustrate this with an Internal BST example. 
          Consider the following internal BST shown and assume a thread is about to delete the node
          \textit{l}. The inorder successor of \textit{l} is \textit{lrl}. It is clear that the delete operation should lock all the nodes on the path from \textit{l} to \textit{lrl} (inclusive). However, this locking scheme is inadequate although it modifies the data structure in a consistent manner. The problem lies outside the code of delete operation itself. The problem surfaces with a concurrent membership operation looking for node \textit{lrl}. Due to node (and henceforth key) movement, it is possible for the traversal code to miss $lrl$. Again, this scenario needs to be inferred from the data structure knowledge. Due to arbitrary key-movement, a concurrent membership operation might claim that it does not see a node as part of the data structure when it is in fact part of the structure. We use the RCU framework in such a case.  
   %  \par
\subsection{Proving  correctness of a Concurrent Algorithm} 

Proving that a set of algebraic operations are thread-safe may involve several proof obligations in general \cite{vafeiadis2006proving}. For linearizable data structures, it is sufficient to show that every execution of the insert, delete and membership operations is equivalent to some sequential execution \cite{herlihy2011art}. This implies that the pre/post-condition invariants associated with the sequential algorithm are never violated in any concurrent execution. That is, when a thread is modifying the data structure with respect to an algebraic operation, it is the only agent in the system modifying the necessary fragment of the data structure. A domain expert who does these proofs by hand in practice, identifies all the destructive update steps performed by each operation. Then, she ensures that if the correct shared memory variables are locked, then any potential interference from other operations does not violate the invariants associated with the destructive update steps for the subject operation. Our reasoner would perform these proof obligations in a way a domain expert would, given the same data structure knowledge and representation. 

\section{Transform Sequential Data Structures to Concurrent Data Structures}

   \subsection{Modeling Thread Interference}
      A concurrency domain expert views interference as arbitrary mutations that can occur on the data structure.  
      We argue that this model is sufficient to discover any undesired thread interactions. The sufficiency of the interference model stems from reasoning interference effects based on sequential algorithm equivalence classes. This feature is usually not present in a concurrent program verification task performed via model checking. However, model checkers may also be instrumented with additional abstractions to guide their search for counterexample traces\cite{vechev2010abstraction}. Our reasoner captures this viewpoint taken by a domain expert and performs the same reasoning.   
     
                       \setlength{\fboxsep}{1pt}

  \begin{center}
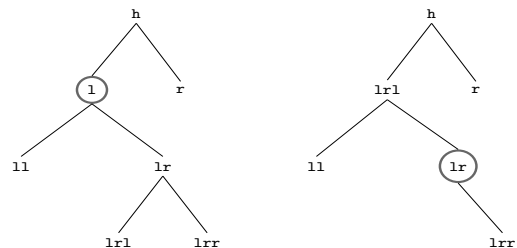

 \resizebox{7cm}{3.3cm}{
\begin{tikzpicture}[
roundnode/.style={circle, draw=black!60,  very thick, minimum size=3mm},
squarednode/.style={rectangle, draw=black!60,  very thick, minimum size=5mm},
-
]
   \fontsize{8}{9}\selectfont\ttfamily
   \tikzstyle{level 1}=[distance=1mm,sibling distance=15mm]
  \tikzstyle{level 2}=[distance=1mm,sibling distance=24mm]
  \tikzstyle{level 3}=[distance=1mm,sibling distance=15mm]
  
  \draw (-5,0) node  {h}
    child {node[roundnode] {l}
             child {node {ll}}
             child {node {lr}
              child {node {lrl}}
              child {node {lrr}}}
             }
    child {node {r}};

    \draw (0,0) node  {h}
    child {node {lrl}
             child {node {ll}}
             child {node[roundnode] {lr}
               child [missing]
               child {node {lrr}}
             }
              }
    child {node {r}};
     
\end{tikzpicture} 
 } 
 \captionof{figure}{Traversal operation reaches till node $l$ but misses $lrl$ by the time it reads $l.right$}
\end{center}
  
    \subsection{Predicate Falsification in Concurrent Execution}
        When designing a concurrent algorithm, it is necessary to know the predicates (invariants) that are falsifiable in a concurrent execution. Since the sequential data structure knowledge provides the necessary preconditions, we systematically check for potential falsification of every conjunct in $pre_{ins}$ (or $pre_{del}$) (Fig. \ref{figure:fig1}) with respect to environment interference. Predicate $pre_{ins}$ (and similarly $pre_{del}$) is defined as $\{pre_{ins}(X,\tau,Y) \leftarrow reach(X), \mathit{suffix(Y)}, edge(X,Y), k_X < k_{\tau} < k_Y\}$ which is picked from the third argument of $pre(ins,block1,..) \in \mathcal{K}$ (Fig. \ref{figure:fig1}). If a predicate is not falsified with respect to the interference model, then it is indeed not falsifiable in \textit{any} serialized concurrent execution. This implies, that a thread need not synchronize on the un-falsifiable predicate(s). For example, in $pre_{ins}$ the predicate $\mathit{suffix(y)}$ is not falsifiable. This is because, any correct algebraic mutation on a linked list (via insert/delete) would only skip the node $y$ but not unlink it in the chain to tail node $t$.  
   \subsection{Lock Acquisition from Critical Conditions} 
         Locks are necessary to protect the invariant predicates from falsification by interference. A conservative approach is to associate locks with every predicate and acquire locks. This approach can be taken for general concurrent programs where less semantic knowledge is available about the sequential program that is being transformed \cite{deshmukh2010logical}. In a fine-grained locking scheme, it is desirable to acquire locks on a precise set of reachable nodes of the pointer data structure.  Intuitively, locking the nodes involved in the window of modification seems sufficient. Although locking this set of nodes is insufficient in general, this sets a lower bound on the number of locks to acquire in a fine-grained locking scheme. Once, the right set of nodes to be locked is guessed, the domain expert confirms the non-falsifiability of the invariants. If non-falsifiability is proven affirmatively, then the concurrent program need only acquire the guessed locks.

  \section{Decomposition of Concurrency Proof Obligations into Reasoning Tasks}
   \subsection{Notations and Assumptions}
    In the following, we define several theories, each of which captures the reasoning tasks performed by a domain expert. Every theory is an ASP program. 
    \begin{enumerate}
    \item We assume the theory $\mathcal{T}_{\mathcal{D}}$ encodes the structural definition of some tree-based recursive data structure $\mathcal{D}$. It also contains various primitives and definitions for every predicate from sequential data structure knowledge. It is also assumed that we can identify predicates that are time invariant from the predicates that are time dependent using $\mathcal{K}$ 
    \item Interference effects of algebraic operations are modelled in the theory $\mathcal{I} \cup \mathcal{T}^\mathcal{R}$, where $\mathcal{I}$ contains all the rules in $\mathcal{T}$ but reified into the time domain. 
    \item We check the adequacy of guessed locks in the theory $\mathcal{I}^\mathcal{L} \cup \mathcal{T}^\mathcal{R}$, where $\mathcal{I}^\mathcal{L}$ is same as $\mathcal{I}$ but considers locking of nodes.
    \item To find the right order of program steps we use the theory $\mathcal{T}^{po}$
    \end{enumerate}
   \subsection{Generating Interference Model}   
        Theory $\mathcal{T}^{\mathcal{R}}$ is the planning domain \cite{lifschitz2019answer} with reified time argument. The theory $\mathcal{T}^{\mathcal{R}}$ contains all the predicates that are time dependent with an extra argument for time. More precisely, for all $p(\bar{X}) \in \mathcal{T}$ that is time dependent, $p(\bar{X}, T) \in \mathcal{T}^{\mathcal{R}}$.
       Also, the ordering of time is captured by the $next \in \mathcal{T}^{\mathcal{R}}$ relation, where $next(t, t')$ means time step $t'$ follows after $t$. 
       
        From the procedural information in $\mathcal{K}$, it is easy to model the instantaneous effects of the actions. We denote the theory encoding the interference model by $\mathcal{I}$. For operation $ins$, a predicate $\mathit{interfere(ins,X,\tau,Y)}$ is added to $\mathcal{I}$ as an abducible. Abducibles are literals that are guessed in ASP.  Note that $pre\_{ins}$ contains exactly the same terms in the data structure procedural knowledge replaced with uppercase variables.  \\
        \par\noindent
        {
         \fbox{
         \begin{minipage}{0.97\linewidth}
         \vspace*{-\bigskipamount}
         \begin{align*}
         \mathit{interfere(ins,X,\tau,Y,T)} \leftarrow  & ~pre\_ins(X,Y,\tau,T), not \ \mathit{neg\_interfere(ins,X,\tau,Y,T)} \\ 
         \mathit{neg\_interfere(ins,X,\tau,Y,T)} \leftarrow &  ~pre\_ins(X,Y,\tau,T), \mathit{not \ interfere(ins,X,\tau,Y,T)} \\
         \mathit{interfere}(del,X,\tau,Y,T) \leftarrow & ~pre\_del(X,Y,\tau,T), \mathit{not \ neg\_interfere(del,X,\tau,Y,T)} \\ 
         \mathit{neg\_interfere}(del,X,\tau,Y,T) \leftarrow & ~pre\_del(X,Y,\tau,T), \mathit{not \ interfere(del,X,\tau,Y,T)}
         \end{align*}
         \end{minipage}
         }
         }
         \par\noindent
         \\
         The causal effects of insert and delete, are also encoded as literals following from \textit{interfere}. They are shown below: 
         \par\noindent \\
         \fbox{
         \begin{minipage}{0.97\linewidth}
         \vspace*{-\bigskipamount}
         \begin{align*}
         edge(X, \tau,T') \leftarrow & ~\mathit{interfere}(ins, X, \tau, Y,T), next(T, T')\\
         edge(\tau, Y,T') \leftarrow & ~\mathit{interfere}(ins,X,\tau, Y, T), next(T, T') \\ 
         edge(X, Y, T) \leftarrow & ~\mathit{interfere}(del,X,\tau,Y,T), next(T, T') 
         \end{align*}
         \end{minipage}
         }
        \par\noindent \\
        In general, for any algebraic operation, $\sigma \in \mathcal{K}$ an interference predicate $\mathit{interfere_{\sigma}}$ is added as an abducible along with its causal effects.  
        \subsection{Checking Falsification Predicates (Task 1)}
        Given the data structure knowledge $\mathcal{K}$, one can discharge conditions that check for falsifications of every conjunct in $pre_{\sigma}$. For $pre_{ins}$ of insert operation in linked list, one can generate predicate falsification checks for \textit{reach, suffix,} and \textit{edge}. These are precisely the time dependent relations in $\mathcal{T}$. The predicate \textit{falsify\_reach} checks for falsification of reach in one time step as follow:  
        $\mathit{\{falsify\_reach}  \leftarrow reach(X, T), not \ reach(X, T'), next(T, T')\}$.
        
        Similarly the falsification of \textit{suffix} and \textit{edge} are defined. These falsification predicates are also added to $\mathcal{I}$. 
        Once the falsification predicates are added to the theory $\mathcal{I}$, one can check if the falsification predicates are true in some model of $\mathcal{T}^{\mathcal{R}} \cup \mathcal{I}$. If their satisfiability is  affirmative, then interference indeed falsifies the predicates. Otherwise, the interference cannot falsify the predicates. This task is an optimization step in order to reduce the number of predicates to be validated post lock-acquisition.
        \subsection{Checking Adequacy of Guessed Locks (Task 2)}
          From the procedural knowledge of the data structure operations, it is easy to guess the locks to be acquired. As an initial guess, every thread should at least synchronize on the nodes involved in the ``window" of modification. For example, the window for insert w.r.t $pre\_ins$ is the set of nodes $\{x,y\}$.
            After guessing the set of locks to be acquired, one can now check their adequacy in the presence of interference. In the interference model, the effects of \textit{interfere} predicates are enabled only if there are no locks already acquired on the nodes they modify. For instance, for $\mathit{interfere}(ins,X,\tau,Y)$  the nodes that are modified are $\{X, \tau\}$ (for arbitrary $X$). Both the two effects shown previously, are enabled only when there are no locks on $X$ or $\tau$. These re-written rules are part of the theory $\mathcal{I}^{\mathcal{L}}$ which represent the reified interference model in the presence of locks. The re-written rules are shown below: 
           \par\noindent \\
             \fbox{
         \begin{minipage}{0.97\linewidth}
         \vspace*{-\bigskipamount}
         \begin{align*}
         edge(X, \tau,T')  \leftarrow & ~\mathit{interfere}(ins, X, \tau, Y,T), next(T, T'),   not \ locked(X, T) \\
         edge(\tau, Y,T')  \leftarrow & ~\mathit{interfere}(ins,X,\tau, Y, T), next(T, T'), not \ locked(\tau,T) \\ 
         edge(X,Y,T')  \leftarrow & ~\mathit{interfere}(del, X, \tau, Y, T), next(T, T'),   not \ locked(X, T)
         \end{align*}
         \end{minipage}
         } 
         
         \par\noindent
         \\
           The locked nodes themselves are captured by the \textit{locked} relation and are added as rules to $\mathcal{I}^{\mathcal{L}}$. The falsification predicates remain the same in $\mathcal{I}^{\mathcal{L}}$. From $\mathcal{I}^{\mathcal{L}}$ , one can infer entailment of the falsification predicates. If falsification is affirmative, then the locking scheme is clearly inadequate. Otherwise, the locking scheme is adequate and the concurrent code can be generated (with lazy synchronization). If the locking scheme is inadequate, we can recommend the RCU framework. 
          \subsection{Validating Sequential Program Order (Task 3)}
              We denote $pre(\sigma)$ and $post(\sigma)$ to be the pre-condition and post-condition associated with operation $\sigma$.
              Given the sequential program steps in $\mathcal{K}$, one should also be able to infer the right order of program steps, that do not violate an invariant, in a concurrent setting. A common invariant that needs to be satisfied is the well-formedness of the data structure. Having the data structure well-formed at all times is desirable as it makes the results returned by membership queries easier to explain with respect to linearizability. Given an invariant $Inv \in \mathcal{T}^\mathcal{D}$, theory $\mathcal{T}^{\mathcal{R}}$ and procedural knowledge in an operation $\sigma$ in knowledge base $\mathcal{K}$, a new theory $\mathcal{T}^{po}$ can be generated that validates the program order of all basic blocks with respect to invariant $Inv$. For every program step $s_{\sigma}(\bar{X}) \in K$, a reified abducible is $s_{\sigma}(\bar{X}, T)$ is generated and added to $\mathcal{T}^{po}$. Then, the $post(\sigma)$ is reified and added to $\mathcal{T}^{po}$. Similarly $Inv$ is also added to $\mathcal{T}^{po}$. Also, the necessary time steps along with their ordering using $next$ is added to $\mathcal{T}^{po}$. Now, if $\mathcal{T}^{po}$ is satisfiable, then there exists a program order that does not violate the $Inv$. The order might be a permutation of the input program order.    
              On the contrary, if $\mathcal{T}^{po}$ is unsatisfiable, then no permutation (including the original program order) exists that can preserve $Inv$. In that case, the reasoner would recommend using an RCU Synchronization. 
             
          \subsection{Detecting Key-Movement (Task 4)}
            As illustrated before for internal BSTs, if the membership operation can potentially miss nodes that are part of the data structure, we use an RCU synchronization. To detect key movement, we compare the sets of keys observed by an asynchronous observer versus a synchronous observer. The asynchronous observer visits one node at a time whereas the synchronous observer visits all the nodes traversable from the beginning node, at every time step. To model this scenario, we require the $next\_node/4$ relation to be present in the knowledge $\mathcal{K}$. The relation $next\_node(x,y,target,t)$ specifies that $y$ is the next node to visit while having last visited $x$ with respect to the $target$ node at time $t$. For no loss of generality on the termination condition of traversal code, we assume that the traversal ends upon reaching the end node. The fact $end\_node(x) \in \mathcal{K}$ states that node $x$ is the end node for the data structure. 
            Key movement is affirmative if there exists a node (key)  that is visited by a synchronous observer, but is not visited by the asynchronous observer at the end of its traversal. Both the rules for synchronous and asynchronous observer use the $next\_node/3$ relation, their definitions given below are straightforward. Finally the predicate $key\_move$ detects key movement as just explained. 
            \par\noindent \\
            \fbox{
            \begin{minipage}{0.97\linewidth}
            \vspace*{-\bigskipamount}
            \begin{align*}
            \hspace*{-1.5em} sync\_visit(X,T) \leftarrow & ~sync\_visit(Y, T), next\_node(Y,X,target,T) \\ 
            \hspace*{-1.5em}  async\_visit(X,T') \leftarrow & ~async\_visit(Y,T), next\_node(Y,X,target,T'),  next\_time(T, T') \\
            key\_move \leftarrow & ~end\_node(X), async\_visit(X, T), sync\_visit(Y, T), not \  async\_visit(Y, T'), T' < T
             \end{align*}
             \end{minipage}

            }
            \\
 \section{Overall Procedure and Soundness}
     Our reasoner performs the above four tasks based on a given data structure theory $\mathcal{T}$ and sequential data structure knowledge $\mathcal{K}$ and takes appropriate decisions on the structure of transformed concurrent code. It is also assumed that $\mathcal{K}$ contains the library of sequential data structure operations $\Sigma = \{\sigma_1, \sigma_2, .. \}$, where each $\sigma_i : S_{\mathcal{D}} \rightarrow S_{\mathcal{D}_{\bot}}$\footnote{$\bot$ signifies that $\sigma_i$ may not be applicable to all instances in $S_{\mathcal{D}}$} is mapping from one instance of data structure $\mathcal{D}$ to another. Without loss of generality we can assume $\Sigma = \{\sigma_1, \sigma_2\}$. We say that the operation $\sigma_i$ is applicable on an instance $\mathcal{D} \in S_{\mathcal{D}}$ if $pre(\sigma_1)$ is true in some model of $\mathcal{T}^\mathcal{R} \cup \mathcal{D}$. There exists a least $\delta \in S_{\mathcal{D}}$ such that each $\sigma_i \in \Sigma$ is applicable to $\delta$. This structure is assumed to be part of $\mathcal{T}^{\mathcal{R}}$. 
     The instance $\delta$ is both necessary and sufficient for the reasoning tasks performed in this paper. It is used in the soundness proof of the procedure later. The intuition behind choosing such an instance $\delta$ is that we need to model executions in which simultaneous operations contend to modify the data structure. If for some $\mathcal{D}'$ there is some $\sigma_i$ that is not applicable to $\mathcal{D}'$ then interference model $\mathcal{I}$ cannot model serialized concurrent execution faithfully, as there might be only a subset of operations modifying the data structure simultaneously and we miss out potential interference that could happen on $\delta$.  A safe concurrent algorithm must take into account interference effects from all destructive update operations in $\Sigma$. Few notations need their description, $pre(\sigma)$ denotes the precondition of some operation $\sigma$, $\mathit{falsify\_p}$ denotes the generated falsification predicate for fluent $p \in \mathcal{T}$, $Locks(\sigma)$ are the set of locks guessed according to some domain expert provided heuristic $\mathcal{H}$ on  $pre(\sigma)$, $Locks\_Adequate$ function checks the adequacy of guessed locks, $Program\_Order$ is the set of all valid program order permutations that preserve a given invariant $Inv(\bar{X})$ and finally, $KeyMove$ is the function that captures the presence of key-movement using similar predicates presented earlier. When the procedure recommends RCU for $\sigma$, then either key-movement is detected or an invariant is violated with any program order $\pi(\sigma)$. 
     \mycomment{If the locks guessed by $\mathcal{H}$ are inadequate, then the user of our system can provide his/her own heuristic $\mathcal{H'}$ and retry.} 
     \[
     \begin{array}{@{}r@{~}c@{~}l@{}}
     \mathit{Unfalsify}  & =  & \{p(\bar{X}) : p \in \mathit{Fluents} \ \land  (\mathcal{T}^\mathcal{R} \cup \mathcal{I} \cup \{\mathit{falsify\_p/0}\}) \models \neg p(\bar{X},T) \} \\
     Locks_{\sigma} & =  & \{\mathcal{H}(pre(\sigma))\}, \text{ where } \mathcal{H} \text{ is the heuristic that guesses locks} \\
     Locks\_Adequate(Locks_{\sigma})  &  = & \left\{\begin{array}{r@{,\quad}l}
                                          \mathit{true} & (\mathcal{I}^L \cup Locks_{\sigma}) \models \neg pre(\sigma) \\ 
                                          \mathit{false} &  \text{otherwise}
                                         \end{array}\right. \\
     Program\_Order(\sigma) & = &  \{\pi(\sigma) : (\mathcal{T}^{po} \cup \pi(\sigma) \cup Inv(\bar{X})) \text{ is satisfiable} \} \\
    KeyMove & = & \left\{\begin{array}{r@{,\quad}l}
                           \mathit{false} &   
                                 \left(\begin{array}{@{}l@{}} \mathcal{T}^{\mathcal{R}} \cup 
                                      \mathcal{I} \cup  \\ \{sync\_visit/2, 
                                      async\_visit/2, \ key\_move/0\}
                                  \end{array}\right)
                                     \models \neg key\_move  \\  
                           \mathit{true}  & \text{otherwise}
                           \end{array}\right.
     \end{array}
     \]
     Without loss of generality assume that we are trying to transform 2 operations of some tree-based inductive data structure $\mathcal{D}$  into their corresponding concurrent versions. Let $\sigma_1, \sigma_2$ denote the two operations. Again, without loss of generality that both $\sigma_1$ and $\sigma_2$ have a single basic block in their destructive update code. For External BSTs insert operation. there are four different pre-conditions and hence four basic blocks. But, as we show, the argument follows similarly if we consider single basic block. Because, $\mathcal{D}$ is inductive, we assume  that $\sigma_1$, $\sigma_2$ are applicable to countably infinite instances of in $S_{\mathcal{D}}$. Clearly, there exists a least instance $\delta \in S_{\mathcal{D}}$ such that $\sigma_1$ and $\sigma_2$ are applicable to $\delta$.  The agents (including interference) that perform $\sigma_1$ or $\sigma_2$ are always cautious with respect to their (permuted) sequential steps from $\mathcal{K}$. That is, after acquiring the desired locks, the agents post-check (validate) their respective preconditions $pre(\sigma_1)$ or $pre(\sigma_2)$ to ensure that the ``window" of modification is still intact and not modified in the time taken to acquire the locks.

    \begin{figure}[h]
    {
    \begin{minipage}{0.7\linewidth}
    \centering
    \begin{algorithmic}
    \scriptsize
    \Procedure{GenerateConcurrentCode}{$\sigma$}
    \State Code $\gets$ $\emptyset$
    \If{$Program\_Order(\sigma) = \emptyset$}
    \State {\textsc{RecommendRCU($\sigma$)}}
    \State \Return
    \EndIf
    \If{$KeyMove(\sigma) = true$}
    \State \textsc{RecommendRCU($\sigma$)}
    \State \Return
    \EndIf
    \If{$Locks\_Adequate(Locks\_{\sigma}) = \mathit{false}$}
    \State \textsc{RecommendRCU($\sigma$)}
    \State \Return
    \EndIf
    \State Code $\gets$ Code $\oplus$ \textsc{LockStmts($Locks_{\sigma}$)}
    \State Code $\gets$ Code $\oplus$ \textsc{Validate($pre(\sigma) \setminus \mathit{Unfalsify}$)}
    \State Code $\gets$ Code $\oplus$ $Program\_Order(\sigma)$
    \State Code $\gets$ Code $\oplus$ \textsc{UnlockStmts($Locks_{\sigma}$)}
    \EndProcedure
    \end{algorithmic}
    \end{minipage}
    } 
    \end{figure}
 
  \subsection{Soundness of Unfalsifiable Predicates and Lock Adequacy Argument} 
   
\noindent \textit{Lemma 1:} If a time-dependent predicate $p(\bar{X}, T)$ (fluent) is unfalsifiable in $\mathcal{I}$ for some $\delta$ where every $pre(\sigma)$ is applicable, then $p(\bar{X}, T)$ is unfalsifiable in any serialized concurrent execution of $\sigma_1$ and $\sigma_2$ \\ 
   \textit{Proof:}  $p(\bar{X})$ may belong to $pre(\sigma_1)$ or $pre(\sigma_2)$ (or both). We denote $\sigma_i$ to mean one of either $\sigma_1$ or $\sigma_2$. \\
   \textit{Case 1:}  $\sigma_i(\delta) = \delta'$ and every $\sigma_i$ is applicable to $\delta'$. Then we have no problem. 
    As conjuncts of $\sigma_1, \sigma_2$ are not falsified including $p(\bar{X})$. \\
   \textit{Case 2:}  $\sigma_i(\delta)$ = $\delta'$ and some $\sigma_j$ is not applicable to $\delta'$. If $p(\bar{X}) \in pre(\sigma_{j'}), j' \neq j$, then we have no problem. If otherwise, $p(\bar{X}) \in pre(\sigma_j)$, there must exist another predicate $p'(\bar{Y}) \in pre(\sigma_j)$ such that $p'(\bar{Y})$ is falsified. Otherwise, $\sigma_j$ would be applicable to $\delta'$ (as $p(\bar{X})$ is not falsified).
   From the above two cases, it is clear that any serialized run of operations $\sigma_1$ and $\sigma_2$ does not falsify $p(\bar{X})$.  
   
   \noindent \textit{Lemma 2:} If the guessed locks for some $\sigma_j$, on an instance $\delta$ where every $pre(\sigma)$ is applicable,  make $pre(\sigma_j)$ unfalsifiable in $\mathcal{I}^\mathcal{L}$, then $pre(\sigma_j)$ is unfalsifiable in any serialized execution of $\sigma_1$ and $\sigma_2$.
   \textit{Proof::} Similar to Lemma 1.
   
  \section{\textit{Locksynth}: A Tool for Automatic Concurrency Synthesis}
 Our tool \textit{Locksynth}\cite{github}, implements the above reasoning tasks in SWI-Prolog and ASP. SWI-Prolog acts as the front-end of the tool that takes the sequential data structure knowledge as input. The backend of our tool is driven by ASP which performs the reasoning tasks discussed above. We use the Clingo ASP Solver. \textit{Locksynth} takes the data structure defitions and sequential code as input and generates the abstract concurrent code. The template for sequential code is as follows: \texttt{code(Operation, BasicBlockNum, Precondition, BasicBlockSteps, PostCondition)}. Here, \texttt{Operation} represents the name of the algebraic operation (ie. insert/delete), \texttt{BasicBlockNum} identifies the basic block, the other arguments carry the same denotation as in \ref{figure:fig1}. In case of Linked Lists, the sequential code is represented as follows:
\begin{lstlisting}[basicstyle=\ttfamily]
code(insert, block1,  
  [reach(x), edge(x,y), key(x,kx), key(y,ky), key(target,ktarget), 
   kx < ktarget, ktarget < ky, not(reach(target)], 
  [link(x, target), link(target, y)], [reach(target)]).
        
code(delete, block1, 
  [reach(x), edge(x,target), edge(target,y), key(x,kx), key(y,ky),
   key(target, ktarget), kx < ktarget, ktarget < ky],
  [link(x,y)], [not(reach(target)]).
        \end{lstlisting}
   We now mention some important features that are part of the implementation. 
 \subsection{Generation of Equivalence Classes via Meta-Interpretation}
   \textit{Locksynth} performs meta-interpretation of the recursive data structure definition. It systematically unfolds data structure instances of increasing size starting from the smallest instance. In case of linked list, \textit{Locksynth} generates the empty list, then generates a list containing one element, two elements and so on. This unfolding is performed to find data structure instances where both insert and delete are applicable. In other words, \textit{Locksynth} finds the equivalence class instances such that both the preconditions of insert and delete are satisfied. The necessity for such an equivalence class instance is already explained in the previous section.  The data structure definition is represented as follows:
  \begin{lstlisting}[basicstyle=\ttfamily]
rule(list,  
     [node(h),key(h,kh),edge(h,X),key(X,KX),lt(kh,KX),suffix(X)]).
rule(suffix(t), []).
rule(suffix(X), 
     [node(X),node(Y),edge(X,Y),key(X,KX),
      key(Y,KY),lt(KX,KY),suffix(Y)]).
  \end{lstlisting}
  
 Within our tool implementation, the query \texttt{?- unfold(Depth, Instance)} generates data structure instances (through meta-interpretation of data structure definition) for a certain given depth. Then the full query, \texttt{?- unfold(Depth, Instances), code(insert, block1, Pre, \_, \_), check(Instance, Pre)} checks if the precondition \texttt{Pre} is satisfied by the data structure instance \texttt{Instance}. The predicate \texttt{check} internally calls the CLINGO ASP solver to perform the satisfiability check, whether the generated \texttt{Instance} indeed satisfies \texttt{Pre}.
 \subsection{Symbolic Reasoning over Nodes, Keys}
   Nodes and Keys are treated as abstract symbols with their usual reflexive and transitive equality and order relations. Symmetry is added in case of equality. This allows the construction of the Data structure instance at a symbolic level. Equality relation over nodes  enables us to separate nodes that are part of the data structure (reachable nodes) from the nodes that are not part of the data structure (unreachable nodes). To illustrate, for the linked list insert operation, we require atleast one node to be not part of the data structure. Therefore, equality reasoning over nodes will allow us to specify a model with an unreachable node, which is not equal to any of the reachable nodes. We represent \texttt{eq\_node(X,Y)} for node equality, \texttt{eq\_num(X, Y)} for key equality and \texttt{lt(X, Y)} for key ordering. The sequential program capturing key order/equality and node equality should use these abstract predicates.   
 \subsection{Reification of predicates into Time Domain and Bounded Time Chains}
    To simulate interference effects, to check for lock adequacy or to detect key movement, we require reasoning over time. To perform temporal reasoning, we reify all fluents into time domain. We also generate the commonsense law of inertia rules handle the frame problem. These rules are similar to the frame rules used in ASP planning problems. A consequence of introducing time is to set the maximum time steps, and link the time steps from the initial time step in a linear chain. This time chain is bounded to the maximum of either the largest basic block (or) the depth of the equivalence class data structure instance. This is because, modelling interference requires only two time steps as interference mutates the data structure atomically. Checking lock adequacy requires also requires two time steps, ie. we simply check if interference has falsified any predicates in the presence of locking. Checking a valid program order requires time steps in the size of largest basic block. Finally, modelling traversal to detect key movement requires a time in the size of the data structure depth. Hence we choose the maximum of either the largest basic block or the data structure depth. For  time chain of length 3, the following facts are added to the appropriate ASP program performing the reasoning task: 
    \texttt{time(t1). time(t2). time(t3). next\_time(t1, t2). next\_time(t2, t3).} The  relation \texttt{next\_time}, establishes the linear order of \texttt{t1, t2} and \texttt{t3}. 
    \subsection{Guessing Locks}
    Given preconditions for insert/delete, a simple heuristic that \textit{Locksynth} follows is to lock every node present in the precondition. These nodes are treated as locked, and are subsequently checked for adequacy in the presence of interference. Locking rules (generated automatically) for guessed nodes for linked list insert are shown below:
    
    \begin{lstlisting}[basicstyle=\ttfamily]
locked(X) :- 
  reach,edge(X, Y),not reach(target),key(X, Kx),key(Y, Ky),
  key(target, ktarget),lt(Kx, ktarget),lt(ktarget, Ky).           
locked(Y) :- 
  reach(X),edge(X,Y),not reach(target),key(X, Kx),key(Y, Ky),
  key(target, ktarget),lt(Kx, ktarget),lt(ktarget, Ky). 
    \end{lstlisting}
 Our locking heuristic locks all the nodes that are present in the operation precondition. Improving this crude heuristic toward a more general lock detection algorithm is part of our future work. 
 \section{Experiments, Conclusion and Future Work}
     Our approach has been applied to Linked Lists, External BSTs and Internal BSTs (Table 1). We are able to synthesize the concurrent versions of insert, delete for Linked Lists and External BSTs. \textit{Locksynth} can also recommend RCU framework for Internal BSTs due to key-movement missed by an asynchronous observer. The generated code for concurrent linked list is shown below:
     \\ \\
    {\small
    \fbox{
    \begin{minipage}{0.45\linewidth}
    \begin{tabbing} aaa \= aaa \= aaa \kill
    \textbf{Sequential Insert}  \\
    \> \{ reach(x), edge(x,y), kx $<$ ktarget, ktarget $<$ ky\}  \\      
    \> link(x, target)   \\ 
    \> link(target, y) \\
    \> \{reach(target)\}    \\ 
    ~  \\ 
    \textbf{Concurrent Insert}  \\
    \> \{reach(x), edge(x,y), kx $<$ ktarget, ktarget $<$ ky\}  \\
    \> lock(x)  \\
    \> lock(y)  \\
    \> lock(target)  \\
    \> if validate(reach(x) \& edge(x,y) \& \\ \>\> ~~~~~~~~~~~kx $<$ ktarget \&  ktarget $<$ ky)\{  \\ 
    \>\> link(target, y)   \\ 
    \>\> link(x, target)  \\
    \> \}  \\
    \> unlock(target)  \\
    \> unlock(y) \\
    \> unlock(x) \\
    \> \{reach(target)\}
    \end{tabbing}
    \end{minipage}
    }
    }
    {\small
    \fbox{
     \begin{minipage}{0.45\linewidth}
     \begin{tabbing} aaa \= aaa \= aaa \= aaa \kill
     \textbf{Sequential Delete} \\
     \> \{reach(x), edge(x, target), edge(target, y), \\ \> ~~kx $<$ ktarget, ktarget $<$ ky\} \\      
     \> link(x, y) \\ 
    ~\> \{not reach(target)\} \\ 
    \textbf{Concurrent Delete}  \\
     \> \{reach(x), edge(x, target), edge(target, y), \\ \> ~~kx $<$ ktarget, ktarget $<$ ky\} \\
    \> lock(x) \\
    \> lock(target) \\
    \> lock(y) \\
    \> if validate(reach(x) \& edge(x,target)  \& \\ \>\>\> ~~~~~edge(target,y) \&   kx $<$ ktarget \& \\
    \>\>\> ~~~~~ktarget $<$ ky)\{ \\ 
    \>\> link(x,y) \\ 
    \> \} \\
    \> unlock(target) \\
    \> unlock(y) \\
    \> unlock(x) \\
    \> \{not reach(target)\} 
    \end{tabbing}
    \end{minipage}
    } 
    }
     \\ \\
      Our work presents the first step towards using commonsense reasoning to generate concurrent programs from sequential data structure knowledge. We have presented the challenges involved in the concurrent code generation and mechanized the reasoning tasks as performed by a human concurrency expert. The procedure described in this paper conforms to McCarthy's vision of building programs that have commonsense and manipulate formulas in first order logic \cite{mccarthy1960programs}.  Our future work aims to apply our technique to more data structures such as Red-Black Trees and AVL-Trees. In general, given the knowledge about a sequential data structure as well knowledge about the concept of concurrency, one should be able to generate suitable, correct versions of concurrent programs. We eventually aim to generalize our technique to arbitrary data structures. Further, the only synchronization primitives we have addressed in this paper are \textit{locks}. However, there are more sophisticated atomic write instructions supported by modern multiprocessors such as \textit{Compare-and-Swap} \cite{valois1995lock}, \textit{Fetch-and-Add} \cite{heidelberger1990parallel}. These atomic instructions give rise to lock-free data structures. Supporting these primitives is part of future work.  
{
\begin{table}
\centering
\begin{tabular}{|c|c|c|c|}
     \hline
      \textbf{Data Structures} &  \textbf{Membership} & \textbf{Insert} & \textbf{Delete} \\
      \hline
      Linked List & No change &  Success &  Success \\
      \hline
      External BST & No change & Success & Success \\
      \hline
      Internal BST & No change & Success & RCU \\ 
      \hline
     \end{tabular}
\caption{Locksynth results for Linked List, Internal and External BSTs}
  \label{tab:Results}
  \end{table}
 } 
   \printbibliography

\end{document}